\documentclass{aa}

\def\be{\begin{equation}}
\def\ee{\end{equation}}
\def\bea{\begin{eqnarray}}
\def\eea{\end{eqnarray}}
\def\l{\lambda}
\def\mkm{\mu{\rm m}}

\usepackage{graphics}
\begin{document}

\thesaurus{08         
           (02.18.6;  
            09.04.1;  
            13.09.3)  
          }

\title{The temperature of non-spherical interstellar
       grains}

\author{N.V.~Voshchinnikov\inst{1},
        D.A.~Semenov\inst{1}
        and
        Th.~Henning\inst{2}}

\institute{Sobolev Astronomical Institute, St.~Petersburg University,
            Bibliotechnaya pl.~2,
            St.~Petersburg--Peterhof, 198904 Russia
\and Astrophysical Institute and University Observatory,
           Friedrich Schiller University,
           Schillerg\"a{\ss}chen 3, D-07745 Jena, Germany}

\offprints{Nikolai V. Voshchinnikov, e--mail: nvv@aispbu.spb.su}
\date{Received $<$date$>$ / Accepted $<$date$>$}
\maketitle
\markboth{N.V.~Voshchinnikov et al.:
 Temperature of non-spherical interstellar  grains}{}

\begin{abstract}
A model of spheroidal particles is used to
 calculate the steady-state temperature of dust grains
immersed in the interstellar radiation field.
It is found that the temperature of
non-spherical grains with the aspect ratios $a/b \la 2$
deviates from that of spheres less than 10\,\%.
More elongated or flattened particles are usually
 cooler than spheres of the same mass and
in some cases the temperatures may differ by even about a factor of 2.
The shape effects
increase with the infrared absorptivity of the grain material
and seem to be more important in dark interstellar clouds.
\end{abstract}

\keywords{Radiation mechanisms: thermal ---
          ISM: dust, extinction ---
          Infrared: ISM: continuum}

\section{Introduction}
The temperature of interstellar dust grains has been
calculated by various authors many times,
starting in the 1940s (see discussion in van de Hulst~\cite{vdh49}).
In the calculations one always assumed that the particles were
spheres (see, e.g., Mathis et al.~\cite{mmp83}).
However, it is well known since
the discovery of interstellar polarization
(Hiltner~\cite{hi49}; Hall~\cite{hall49}; Dombrovski~\cite{d49})
that the non-spherical grains should exist in the interstellar medium.

The first and  single attempt to study the shape effects on
grain temperature was made by Greenberg \& Shah~(\cite{gs71}).
They considered metallic and dielectric
Ray\-leigh spheroids and infinitely long icy cylinders of radius
0.1 $\mu$m. Their conclusion that non-spherical particles
are about 10\,\% cooler than spheres is the result
even included in textbooks (see Whittet~\cite{w92}).

The calculation of the dust temperature is an essential step
in any modelling of infrared (IR) emission from dust shells
and discs, interstellar clouds, and galaxies.
The dust temperature is included into the expressions for
the determination of dust mass and cooling processes.
The temperature of particles is also important for the process
of molecule formation on grains.

In this letter, we estimate the particle shape
effects on the interstellar grain temperature.
Calculations are made for compact homogeneous
prolate and oblate spheroids
of different semiaxes ratios, sizes and compositions.

\section{Calculations}
Let us consider an interstellar grain in thermal equilibrium with its
surroundings. In the isotropic radiation field the
grain temperature $T_{\rm d}$ can be obtained as a solution of the
energy balance equation for absorbed and emitted energy [in erg\,s$^{-1}$]
\be
\int^\infty_0 \overline{C}_{\rm abs} \ 4\pi J_{\l}^{\rm ISRF} \,{\rm d}\l =
\int^\infty_0 \overline{C}_{\rm em} \ 4\pi B_{\l}(T_{\rm d}) \,{\rm d}\l\,, \label{eq1}
\ee
where $\overline{C}_{\rm abs}$ and  $\overline{C}_{\rm em}$ are
the absorption and emission cross-sections
averaged over orientation and $4\pi J_{\l}^{\rm ISRF}$
the interstellar
radiation field (ISRF) [in erg\,cm$^{-2}$\,s$^{-1}$\,$\mu$m$^{-1}$].

We suppose that the grains are prolate or oblate homogeneous
spheroids with the aspect ratio $a/b$ ($a$ and $b$ are the major and minor
semiaxes of a spheroid, respectively).

We characterize the particle size by the radius $r_{\rm V}$
of the sphere whose volume is equal to that of a spheroid.
The major semiaxis of the spheroid is connected with $r_{\rm V}$
as follows:
\be
a = r_{\rm V} ({a}/{b})^{2/3}
\ee
for prolate spheroids and
\be
a = r_{\rm V} ({a}/{b})^{1/3}
\ee
for oblate ones.
In our calculations,  particles with sizes
$r_{\rm V} = 0.005 - 0.25\,\mkm$ are considered.

Under interstellar conditions, we can generally assume that
the incident radiation is non-polarized and the grains are
arbitrarily oriented in space (3D-orientation). Then
the mean absorption cross-sections can be found as
%
%
\bea
\overline{C}_{\rm abs} =
\int^{\pi /2}_0 \, \frac{1}{2} \,
\left[{Q}_{\rm abs}^{\rm TM}(m_\lambda,r_{\rm V},\l,a/b,\alpha) + \right. &
 \nonumber \\
\left.
{Q}_{\rm abs}^{\rm TE}(m_\lambda,r_{\rm V},\l,a/b,\alpha)\right]
G(\alpha) \sin\alpha \,{\rm d}\alpha \,. \label{avc} &
\eea
Here, $m_\lambda$ is the refractive index of the grain material,
$\alpha$ the angle between the rotation axis of
the spheroid and the wave-vector ($0\degr \leq \alpha \leq 90\degr$) and
$G$  the geometrical cross-section of a spheroid (the area
of the particle shadow) which is
\be
G(\alpha) = \pi r_{\rm V}^2 \left (\frac{a}{b} \right ) ^{-2/3}
            \left[\left (\frac{a}{b} \right )^{2}\sin^2\alpha
            + \cos^2\alpha\right]^{1/2}
\label{Gp}
\ee
for a prolate  spheroid and
\be
G(\alpha) = \pi r_{\rm V}^2 \left (\frac{a}{b} \right ) ^{2/3}
            \left[\left (\frac{a}{b} \right )^{-2}\sin^2\alpha
            + \cos^2\alpha \right]^{1/2}\,.
\label{Go}
\ee
for an oblate spheroid.

The energy emitted by a particle is proportional to its surface area.
Then the emission cross-sections can be found as
%
%
\bea
\overline{C}_{\rm em} = S
\int^{\pi /2}_0 \, \frac{1}{2} \,
\left[{Q}_{\rm abs}^{\rm TM}(m_\lambda,r_{\rm V},\l,a/b,\alpha) + \right. &
 \nonumber \\
\left.
{Q}_{\rm abs}^{\rm TE}(m_\lambda,r_{\rm V},\l,a/b,\alpha)\right]
 \sin\alpha \,{\rm d}\alpha \,, \label{ave} &
\eea
where
\be
S = 2 \pi r_{\rm V}^2
            \left[\left (\frac{a}{b} \right )^{-2/3}
           + \left (\frac{a}{b} \right )^{1/3}\frac{\arcsin (e)}{e}
           \right]
\label{sp}
\ee
for an prolate  spheroid and
\be
S = 2 \pi r_{\rm V}^2
            \left[\left (\frac{a}{b} \right )^{2/3} +
            \left (\frac{a}{b} \right )^{-4/3}
            \frac{\ln [(1+e)/(1-e)]}{2 e}
           \right]
\label{so}
\ee
for an oblate spheroid and $e=\sqrt{1-(a/b)^{-2}}$.

In Eqs.~(\ref{avc}), (\ref{ave}),
the superscripts TM and TE are related to two cases of
the polarization of incident radiation (TM and TE modes).
The efficiency factors  $Q_{\rm abs}^{\rm TM,TE}$ are calculated from
the solution to the light scattering problem for spheroids
(see Voshchinnikov \& Farafonov~\cite{vf93} for details).
The benchmark results given by Voshchinnikov et al.~(\cite{v99})
were used for a thorough testing of the numerical code.

The chemical composition of interstellar grains is a
subject of continuing discussion. As usual,
a mixture of carbon and silicate particles or
composite grains are considered (see Henning~\cite{h98}
for a recent review).
We consider six species used earlier by Il'in \& Voshchin\-ni\-kov (\cite{iv98})
in the modelling of radiation pressure in envelopes of late-type
giants. They are: an amorphous carbon (AC1), iron and magnetite
as examples of highly absorbing materials;
the astronomical silicate (astrosil), artificial dirty silicate
(Ossenkopf et al.~\cite{ohm92}; OHM-silicate)
and clean glassy pyroxene 
as examples of different types of silicates.
The choice of the optical constants of these materials is
described by Il'in \& Voshchinnikov~(\cite{iv98}).\footnote{The
refractive indices also may be found in the database of optical constants
(Henning et al.~\cite{h99}).}
This sample was extended by two species:
carbon material (cellulose) pyrolized at 1000$\degr$\,C
(cel1000; J\"{a}ger et al.~\cite{j98}) and dirty ice  used
in the classical work of Greenberg \& Shah~(\cite{gs71}).
In the last case, we take
the imaginary part of the refractive index $k = 0.02$
in the wavelength range $\l\l = 0.17 - 1.2 \,\mkm$ as it was
made by Greenberg~(\cite{g68}, \cite{g71}).

The interstellar radiation field is adopted according to
Mathis et al.~(\cite{mmp83}) for the solar neighbourhood
\be
4\pi  J_{\l}^{\rm ISRF} = 4\pi  J_{\l}^{\rm UV} +
4\pi  \sum^3_{j=1} W_j B_{\l}(T_j) \,,
\label{isrf}
\ee
where the UV emission from early type stars $4\pi  J_{\l}^{\rm UV}$
is given by Mezger et al.~(\cite{mmp82}).
Other components of the
ISRF are described by blackbody radiation with the temperatures
$T_1 = 7500$\,K, $T_2 = 4000$\,K, $T_3 = 3000$\,K, respectively.
The corresponding dilution factors are
$W_1 = 10^{-14}$, $W_2 = 10^{-13}$, $W_3 = 4 \, 10^{-13}$.

\section{Results and discussion}
\begin{figure}
\resizebox{8.8cm}{!}{\includegraphics{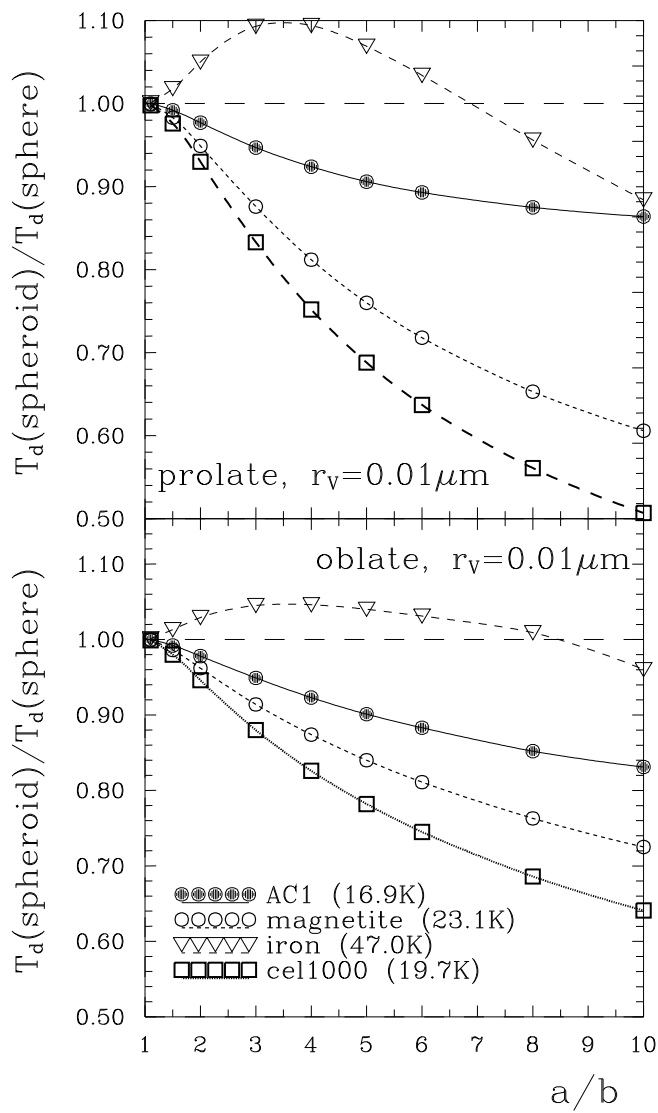}}
\caption[]{The shape dependence of the ratio of temperatures
for prolate (upper panel) and oblate (lower panel)
spheroids and spheres of materials with metallic properties.
The grain composition and the temperature
of spherical particles are indicated.}
\label{f1}
\end{figure}

\begin{figure}
\resizebox{8.8cm}{!}{\includegraphics{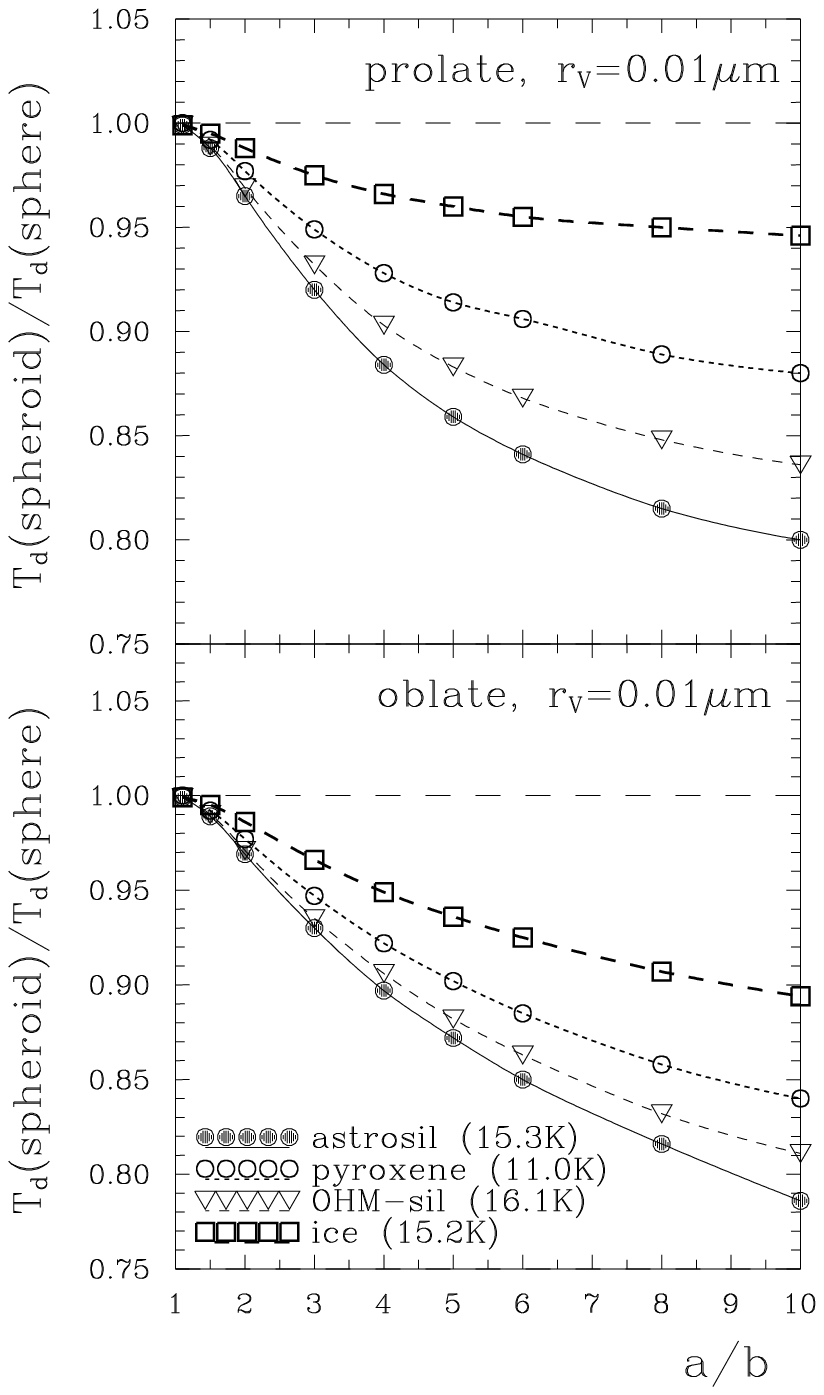}}
\caption[]{The same as Fig.~\ref{f1} but now for dielectric materials.}
\label{f2}
\end{figure}

We  calculated the temperatures of prolate and oblate sphe\-roi\-dal
particles with the aspect ratios $a/b = 1.1 - 10$ and
compared the results with the temperatures of spherical
particles of the same volume (or mass). The results are given
in Table~1 for three basic materials (amorphous carbon,
astrosil and ice) and $a/b = 4$.
They are also shown in Figs.~\ref{f1} and
~\ref{f2} for particles of single size $r_{\rm V} = 0.01 \, \mkm$.
\begin{table*}
\begin{flushleft}
\caption[]{The temperature of spherical and spheroidal ($a/b=4$) grains
in Kelvin}
\begin{tabular}{cccccccccccc}
\noalign{\smallskip}
\hline
\noalign{\smallskip}
& \multicolumn{3}{c}{Amorphous carbon} && \multicolumn{3}{c} {Astronomical silicate}
&& \multicolumn{3}{c} {Ice}\\
\noalign{\smallskip}
\cline{2-4} \cline{6-8} \cline{10-12}
\noalign{\smallskip}
 $r_{\rm V}$,\,$\mkm$
& Sphere  & Prolate & Oblate &
& Sphere  & Prolate & Oblate &
& Sphere  & Prolate & Oblate \\
\noalign{\smallskip}
\hline
\noalign{\smallskip}
 0.005 & 16.8  & 15.6 & 15.6 && 15.1 & 13.4 &13.6 &&15.2 &14.7 &14.4 \\
 0.010 & 16.9  & 15.6 & 15.6 && 15.3 & 13.5 &13.7 &&15.2 &14.7 &14.4 \\
 0.020 & 17.1  & 15.7 & 15.6 && 15.3 & 13.4 &13.5 &&15.2 &14.6 &14.3 \\
 0.030 & 17.1  & 15.7 & 15.6 && 15.1 & 13.2 &13.3 &&14.9 &14.4 &14.2 \\
 0.050 & 17.2  & 15.6 & 15.5 && 14.7 & 12.9 &13.0 &&14.5 &14.1 &13.8 \\
 0.100 & 17.4  & 15.6 & 15.5 && 14.4 & 12.5 &12.2 &&14.0 &13.6 &13.4 \\
 0.150 & 17.4  & 15.4 & 15.3 && 14.4 & 12.1 &12.3 &&13.9 &13.4 &13.2 \\
 0.250 & 16.9  & 14.8 & 14.8 && 14.4 & 12.3 &12.3 &&13.7 &12.8 &12.6 \\
\noalign{\smallskip}
\hline
\end{tabular}
\end{flushleft}
\label{tab1}
\end{table*}

As it is seen from Table~1,
the size effects on the temperature
for non-spherical particles  are rather  small
as it is the case for spherical  grains.


The shape effects are the most prominent in the case of
grains consisting of absorbing materials (see Figs.~\ref{f1}, \ref{f2}).
Such particles emit more radiation at long wavelengths
where the imaginary part of the refractive index $k$
is large and increases significantly with wavelength (excluding
AC1). In comparison with other materials presented in Fig.~\ref{f1},
the iron particles absorb much more radiation at visual wavelengths.
For intermediate values of $a/b$, the emissive capacity
of spheroids is almost the same as for spheres
and therefore spheroids have larger temperatures. This explains the
peculiar curves for iron in Fig.~\ref{f1}.
Note also that for extremely prolate or oblate particles
the decrease of the temperature with the growth $a/b$ ceases.
For example,  for prolate particles from cel1000
the ratio $T_{\rm d}({\rm spheroid})/T_{\rm d}$\\$({\rm sphere})
=$ 0.51, 0.38, 0.28 and  0.26 if $a/b = 10, 20, 50$ and 100,
respectively.

The behaviour of $k$ for all silicates is rather similar
and the values of $k$ in the IR are usually smaller than for
materials with metallic properties. As a result, the ratio
$T_{\rm d}({\rm spheroid})/T_{\rm d}({\rm sphere})$
changes in narrower limits
in Fig.~\ref{f2} in comparison with  Fig.~\ref{f1}.

The behaviour similar to described above is kept
when the incident radiation is slightly polarized ($\leq 10\,\%$)
or its spectral distribution changes due to attenuation of
the UV field and other components in Eq.~(\ref{isrf}).
In both cases the relative changes of the ratio of
temperatures remain in the limits smaller than 5\,\%.

The alignment of dust grains does not affect strongly
the temperature of non-spherical interstellar grains.
Its influence becomes more important for particles
immersed in an anisotropic radiation field like
in circumstellar shells (especially in the case of oblate grains,
see Voshchinnikov \& Semenov~\cite{vs99} for discussion).
A similar situation exists near the the edges of dark interstellar clouds.
For a fixed $r_{\rm V}$ the difference in
temperatures of porous spherical and non-spherical particles is
smaller than for compact ones.

The temperature of interstellar dust grains can be found by fitting
the galactic IR emission by modified blackbody curves.
The dust emission spectrum obtained from COBE data for dust
associated with H\,{\sc I} gas can be represented by a single modified blackbody
curve with $B_{\nu}(17.5\,{\rm K}) \nu^2$
(Boulanger et al.~\cite{b96}).
In order to compare the observa\-ti\-onal\-ly-ba\-sed emissivity law with dust
models, the shape, size, and porosity distribution of the particles have
to be taken into account. It is
not the goal of this paper to perform such an analysis, but to provide
necessary input data. From Table~1,
it can be clearly seen that the temperature of refractory spheroidal
grains with $a/b=4$ ranges between 12.1\,K
and 15.7\,K, lower than predicted by the observations.

The steady-state temperature of grains also
dictates the efficiency of the process of molecule formation
on grain surfaces. As it is shown by  Pirronello et al.~(\cite{p99}),
the decrease of a grain temperature by 20 -- 30\,\%
can enhance the efficiency of hydrogen recombination
by 2 -- 4 times.

\section{Conclusions}
In the conditions typical of  the interstellar radiation field
the temperature of the non-spherical (spheroidal) grains
deviates from that of spheres of the same volume less than 10\,\%
if the aspect ratios $a/b \la 2$.
More elongated or flattened particles are usually
cooler than spheres and
in some cases the temperatures may differ by a factor 2 and more.
The shape effects are almost independent of particle size
but increase with the  growth of the material
absorption in the infrared (i.e., they are more important
for carbonaceous and metallic particles than for silicates and ices).
In dark interstellar clouds
the non-spherical particles will be cooler than spheres,
facilitating the molecule formation  on grain surfaces.

\acknowledgements{The authors are thankful to Vladimir Il'in
and John Mathis, the referee, for  comments.
The work was partly supported by grants of
the Volkswagen Foundation,
the program ``Universities of Russia -- Fundamental Researches''
(grant N~2154) and
the program ``Astronomy'' of the government of the Russian Federation.}

\end{document}